# The role of heavy fermions

Anna Hasenfratz and Thomas A. DeGrand [a]

[a]Department of Physics, University of Colorado Boulder, Colorado 80309

Heavy dynamical fermions with masses around the cut-off do not change the low energy physics apart from a finite renormalization of the gauge coupling. In this paper we study how light the heavy fermions have to be to cause more than this trivial renormalization.

## 1. INTRODUCTION

Heavy fermions in QCD studies are "real": they are present either as real physical effects (heavy quarks) or as the consequence of the regularization (Wilson fermions). In all cases the influence of heavy fermions at low energies is expected to be no more than some induced effective gauge coupling.

When can we expect that the fermions influence the physical spectrum in a non-trivial way and when can we just replace them with an effective local gauge action? The answer obviously depends on the physical processes we are investigating. Heavy fermions are always present in the spectrum, unless their mass is above the cut-off, but if the low lying gauge and light quark hadronic spectrum is much below the energy level of the heavy fermions they will not directly influence the low energy spectrum.

This paper is the summary of Ref. [1] where we addressed the above questions in detail.

## 2. THE INDUCED EFFECTIVE GAUGE ACTION

The fermions' induced gauge coupling can be calculated by evaluating a 1-loop graph if the fermions are heavy [2][3]. Consider the lattice regularized model of $\tilde{N}_f$ fundamental (Wilson) fermions interacting with $SU(3)$ gauge fields, whose action is

$$S = \beta \sum_{n,\mu} Tr(U_p) + \frac{1}{2\kappa} \sum_{n,m} \bar{\psi}_n K_{nm}[U]\psi_m, \quad (1)$$

where

$$K_{nm}[U] = \delta_{nm} - \kappa \sum_\mu ((r - \gamma_\mu) U_{n\mu} \delta_{n+\mu,m} + (r + \gamma_\mu) U_{n\mu}^\dagger \delta_{n-\mu,m}). \quad (2)$$

$r = 1$ corresponds to the usual Wilson fermion formulation while $r = 0$ describes $N_f = 16 \times \tilde{N}_f$ staggered fermions. Integrating out the fermions we obtain the effective gauge action of the form

$$S_{eff} = S_g - Tr \ln K[U]. \quad (3)$$

Using the continuum representation of the gauge field $U_{n\mu} = e^{iagA_\mu(n)}$ one can express $S_{eff}^{ferm}$ in terms of the continuum fields $A_\mu(n)$ as the sum of one loop diagrams. The leading term of the effective action is the usual continuum gauge action $\frac{1}{g_0^2} F_{\mu\nu} F_{\mu\nu}$ where the coefficient $1/g_0^2$ can be calculated by evaluating two 1-loop graphs.

The result is given by a four-dimensional lattice integral

$$\frac{1}{g_0^2} = \frac{\tilde{N}_f}{4} \int \frac{d^4 p}{(2\pi)^4} Tr \left\{ Q(p_\mu) S(p) Q(p_\mu) \frac{\partial^2}{\partial p_\nu^2} S(p) \right\} \quad (4)$$

where $S(p)$ is the lattice fermion propagator $S^{-1}(p) = \frac{1}{2\kappa} - r \sum_\mu \cos(p_\mu) - i \sum_\mu \gamma_\mu \sin(p_\mu)$ and $Q(p_\mu) = ir \sin(p_\mu) + \gamma_\mu \cos(p_\mu)$.

The effective action has additional terms containing more derivatives and/or external gluon legs. These graphs are multiplied by negative powers of $m$ and are suppressed for heavy fermions [2]. In the limit where the higher order terms can be neglected the effective action is indeed a pure gauge action with bare coupling constant given by Eqn. 4. In terms of the plaquette action lattice model it corresponds to an effective plaquette term with coefficient $\Delta\beta = 6/g_0^2$.

When is the above approach valid? The question is two-fold: 1) Can the non-local effective action indeed be replaced by a single plaquette term and 2) how well does Eqn. 4 predict the coefficient of this term? It is possible to have a pure gauge effective action in a region where Eqn. 4 is no longer valid.

## 3. VALIDITY OF THE EFFECTIVE MODEL

We studied the dependence of the finite temperature confinement-deconfinement phase transition on $N_T, N_f$ and $m$.

The quenched phase transition at $N_T = 4$ is at $\beta_c^Q = 5.69(1)$ [4]. Introducing $N_f$ flavors of fermions with mass $m$ will shift the transition to $\beta_c^{N_f} = \beta_c^Q - \Delta\beta$. If $m$ is such that the fermionic action can be considered pure gluonic at low energies then $\Delta\beta(N_f, m) = N_f \Delta\beta_1(m)$. If, in addition, $m$ is large enough that the perturbative formula is valid, $\Delta\beta_1(m)$ is given by Eqn. 4. Thus we expect the following behavior for the shift $\Delta\beta(N_f, m)$: For $m \gg \Lambda$ where the fermion and gluon mass scales are well separated we expect to see universal behavior $\Delta\beta/N_f = f(m)$ where $f(m)$ is given by Eqn. 4. For smaller $m$ we expect Eqn. 4 to fail quantitatively. However, it might happen that $\Delta\beta/N_f$ is still some universal function of the quark mass. Finally, when the fermion scale is the same order as the gauge scale one can no longer replace the fermions by an effective gauge action. The shift $\Delta\beta/N_f$ would then be different for different $N_f$, $N_T$. Measuring the finite temperature transition for different $N_f$ and $m$ values makes it possible to distinguish the different scenarios.

The finite temperature transition is first order for the pure gauge theory, and is stable under the inclusion of heavy fermions. At sufficiently light quark mass the deconfinement transition line terminates. We might still be able to track the crossover point as a function of $N_f$ and $m$, as long as the fermionic spectrum remains heavy compared to the low energy gluon spectrum.

At very small or zero quark mass (depending on the number of light flavors) there is a second transition whose behavior is thought to be primarily chirally-restoring. At this transition the role of the fermions is fundamental and one would not expect the decoupling of the gluonic and fermionic spectrum.

In the above consideration we had to assume the relation $\beta = 6/g_0^2$ - the induced gauge coupling is expressed through the bare continuum coupling $g_0^2$ while in a lattice simulation one uses the coefficient of the plaquette term $\beta$. $\beta = 6/g_0^2$ should hold in the continuum, large $\beta$ limit; one expects to encounter deviations when the finite temperature transition happens in the strong coupling (small $\beta$) region.

In the following we collect the available numerical data on finite temperature phase transition for different $N_T$, $N_f$ and mass values [1]. To compare simulations with different $N_f$ and $N_T$ values we translate numerical data to express the shift in the gauge coupling caused by one of the fermions only, $\Delta\beta_1 = (\beta_c^Q - \beta_c^{N_f})/N_f$. Here $\beta_c^Q$ is the Monte Carlo quenched critical coupling and $\beta_c^{N_f}$ is the Monte Carlo $N_f$ flavor critical coupling.

The results are collected in Fig. 1 where the solid line corresponds to the analytical prediction, Eqn. 4.

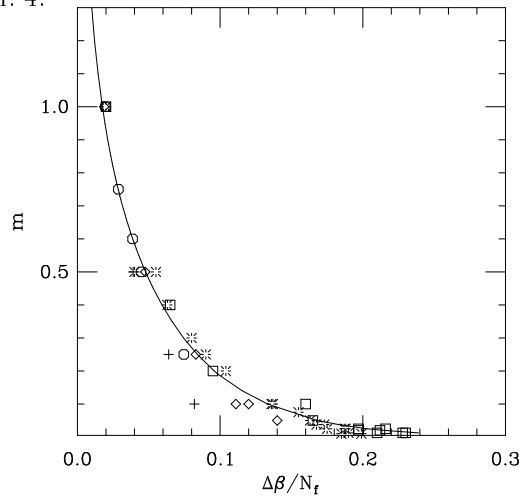

Figure 1. The induced gauge coupling divided by the number of flavors, $\Delta\beta/N_f$, compared with the curve from Eqn. (4) as a function of quark mass. Data are labeled with octagons for $N_f = 24$, plusses for $N_f = 17$, diamonds for $N_f = 8$, bursts for $N_f = 4$, and squares for $N_f = 2$.

$N_f = 24$: The octagons in Fig. 1 are from a 24 flavor staggered fermion simulation on $6^3 \times 4$ lattices. The data points agree with the analytical prediction for $m > 0.25$. At $m = 0.25$ simulations with $N_f \leq 8$ agree with the analytic prediction. The deviation here should be attributed to the fact that $\beta_c = 3.90(5)$ is a very strong coupling where $\beta = 6/g_0^2$ does not hold anymore.

$N_f = 17$: (Plusses) These data are from runs using the Langevin updating algorithm on $N_T = 4$ lattices. The analytic formula consistently overestimates the shift in $\beta_c$. This is hard to understand given that the $N_f = 24$ and $N_f = 8$ simulations (see below) are well represented by the formula. We suspect the deviation is due to integration timestep errors in the simulation.

$N_f = 8$: (Diamonds) These data are also from runs using the Langevin updating algorithm on $N_T = 4$ and 6 lattices. The analytic formula accurately predicts the location of the transition or crossover point for the larger values of the quark mass studied. At very small $m$ and $N_T > 8$ a number of authors [4] have seen a transition which may be a bulk transition. Our analytic formula does not predict this transition. On the other hand at such small mass values Eqn. 4 is not expected to be valid anymore.

$N_f = 4$: (Bursts) These simulations do not show a phase transition at moderate values of the quark mass. At small $m$ they show a first order transition which is believed to be associated with chiral restoration. The location of the transition/crossover is well predicted by Eqn. 4 down to $m = 0.05$. For $N_T = 4$ at $m = 0.073$ the first order chiral transition switches on [4]. Surprisingly, Eqn. 4 is still valid. One can see deviation from the analytic formula for $m \leq 0.025$.

$N_f = 2$: (Squares) Most of the $N_f = 2$ simulations were performed at very light values of the quark mass. They do not show a phase transition; instead, they show a smooth crossover from a chirally broken phase to a chirally restored one. Nevertheless, the location of the crossover point is very well tracked by the analytic formula, even at very light values of the quark mass.

The agreement of the data in Fig. 1 with the analytic prediction, especially with smaller $N_f$, is remarkable even for masses as small as $m = 0.05$ or below. The fact that the data appear to lie on a universal curve is a signal that the fermions induce an effective $\beta$ whose strength is linear in $N_f$ at fixed quark mass, down to very small mass.

## 4. SUMMARY

We demonstrated that the effects of fermions on the finite temperature phase transition can be described by an induced effective plaquette term for masses as low as $m \simeq 0.05$. The induced coupling is proportional to the flavor number and is independent of $N_T$. The proportionality constant is given by a simple 1-loop formula. From the point of view of lattice simulations of QCD, our results show that dynamical quarks must be very light to cause interesting effects. A finite temperature simulation at some quark mass ought to show an induced $\beta$ which is not given by the one-loop formula, before one could claim that a $T = 0$ simulation at the same mass would be sensitive to the effects of dynamical quarks. This is just barely the case in contemporary dynamical fermion simulations.

This work was supported by the National Science Foundation and the Department of Energy.